\documentclass{scrartcl}

\usepackage[sumlimits,intlimits,namelimits]{amsmath}
\usepackage{amssymb}
\usepackage[english]{babel}
\usepackage{bbm}
\usepackage{parskip}

\newcommand{\Z}{\mathbbm{Z}}
\newcommand{\id}{\mathbbm{1}}

\newcommand{\ud}{\text{d}}
\newcommand{\ui}{\text{i}}

\renewcommand{\vec}[1]{\boldsymbol{#1}}

\newcommand{\p}{\partial}

\begin{document}
\bibliographystyle{plain}
\setlength{\parindent}{0cm}

\title {Supersymmetric Branes in Generic M-Theory Attractor Geometries}
\author{S.~Ansari$^{a}$\footnote{\tt{ansari@theorie.physik.uni-muenchen.de}}, J.~D.~L\"ange$^{a}$\footnote{\tt{jdl@theorie.physik.uni-muenchen.de}}, I.~Sachs$^{a,b}$\footnote{\tt{ivo@theorie.physik.uni-muenchen.de}} \\[10pt]
$^{a}$Arnold-Sommerfeld-Center for Theoretical Physics\\
Theresienstra\ss e 37, D-80333 M\"unchen, Germany\\
\\
$^{b}$Department of Physics and Astronomy, UCLA\\
Los Angeles, CA 90095-1547, USA\\}

\date{}

\maketitle

\vspace{-250pt}
\hfill{LMU-ASC 12/07}
\vspace{+250pt}

For a given attractor black hole with generic D6-D4-D2-D0-charges in four dimensions we identify the set of supersymmetric branes, static or stationary in global coordinates, of the corresponding eleven-dimensional  near horizon geometry. The set of these BPS states, which includes branes which partially or fully wrap the horizon, should play a role in understanding the partition function of black holes with D6-charges. 
\clearpage

\section{Introduction}
An outstanding problem in studying extremal black holes in string theory and their relation to conformal field theories is to get a precise microscopic description of  four-dimensional black holes with D6-D4-D2-D0-charges obtained by compactification of type IIA string theory on a Calabi-Yau three-fold $X$. The charges are due to D-branes completely wrapped on non-trival cycles of $X$. For generic charges one expects this black hole geometry to be dual to some conformal quantum mechanics on the boundary of the near horizon $\text{AdS}_2$ geometry. 

For vanishing D6-charge the geometric entropy of such black holes can be given a microscopic 
understanding upon lifting this solution to M-theory. From the M-theory perspective this class of black holes are obtained by wrapping M5-branes with fluxes and momentum along the M-theory circle on a four-cycle in $X$. The corresponding near-horizon geometry is dual to some $1\!+\!1$-dimensional conformal field theory which lives on the dimensionally reduced five-brane world volume \cite{MSW}. This observation allowed the authors of  \cite{MSW}  to derive the asymptotic degeneracy of states using standard methods of conformal field theory. Upon compactification to IIA theory the near horizon geometry obtained is $\text{AdS}_2 \times S^2$. For this model a candidate for a dual quantum mechanics for the D4-D0 black hole has been proposed \cite{StromingerBHQM} in terms of the degrees of freedom of probe D0-branes in this background (see also \cite{Mathur,Aspinwall} for related discussions).

Lifting a black hole solution with D6-charge to M-theory one obtains instead a Calabi-Yau compactification to a Taub-NUT geometry with fluxes \cite{Strominger4D5D}. While for large distances compared to the size of the asymptotic Taub-NUT circle these black hole geometries are effectively four-dimensional, the M-theory circle near the horizon is proportional to the size of the horizon so that the M-theory perspective is more appropriate. Essentially,  the near horizon geometry is a five-dimensional  spinning black hole sitting at the center of the Taub-NUT geometry. This 4D-5D connection has been exploited in \cite{Strominger4D5D} to relate a certain partition function of a class of four-dimensional black holes to that of five-dimensional black holes. The near horizon-geometry in five dimensions is essentially $\text{AdS}_2\times S^3/\Z_{p^0}$. Consequently for non-vanishing D6-charge the problem is not directly related to a $1\!+\!1$-dimensional CFT\footnote{In fact it has been argued in \cite{Diaconescu} that black holes 
with D6-charge are related through a chain of string dualities to BPS states without D6-charge. A similar relation was also conjectured in  \cite{Strominger07} based on an embedding of space-time in the total space of the $U(1)$-gauge bundle over near horizon geometry of the black hole. It would be interesting to see how these two approaches are related.}. 

On the other hand it appears that supersymmetric probe branes in the near horizon geometry play a 
role in understanding the dual quantum mechanics and the black hole partition function. They can be thought of as the ``constituents'' of the black hole in question. The asymptotic degeneracy of the electric consituents in the background flux geometry supported by the magnetic charges accounts for the black hole entropy in some cases \cite{StromingerBHQM,Aspinwall}. For instance it has been shown in \cite{StromingerBHQM} that the ground state degeneracy of D0-branes in a D4-brane flux background reproduces the correct asymptotic degeneracy for D4-D0 charge black holes. 
Here the relevant Hamiltonian is the one of conformal quantum mechanics on the moduli space of D0-branes in the flux background. An important subtlety is though that the appropriate 
Hamiltonian appears to be that which generates translation in ``global'' time rather than Poincar\'e 
time which coincides with asymptotic time\footnote{The Poincare-Hamiltoinian has a contionus spectrum with no ground state as a consequence of the incompletness of classical dynamics in Poincar\'e coordinates \cite{Townsend}.}. The dominant contribution to the entropy comes from D0-branes bound to two-branes wrapping the horizon of the black hole. 
 
On another front it has been shown \cite{StromingerOSV} that the elliptic genus of the $(0,4)$-CFT \cite{MSW} dual to black holes with D4-D2-D0-charge has a dilute gas expansion dominated by multi-particle chiral primaries which are just the stationary M2 (and anti-M2) branes in global $\text{AdS}_3$-coordinates, wrapped on holomorphic curves in the Calabi-Yau and sitting at the center of $\text{AdS}_3$. This provides a derivation of the OSV-conjecture relating the mixed partition function of the black hole to the square of the topological string partition function \cite{OSV}. 

The purpose of the present paper is to prepare the ground for extending the above-mentioned results \cite{StromingerBHQM,StromingerOSV} to black holes with D6-charge by describing the supersymmetric probe branes in D6-charge backgrounds. Of course,  in this case we do not have a known ``parent'' $1+1$-dimensional CFT to compare the probe-brane degeneracies with. Nevertheless one can hope that understanding the degeneracies of these states will give some insight about the underlying microscopic theory. In Section 2 we will describe the eleven-dimensional near horizon geometry of a 4D black hole with generic D6-D4-D2-D0 charge. While the full space-time geometry of a generic black hole with D6-charge  is a solution of five-dimensional $\mathcal{N}=2$ supergravity with $n_v-1$ vector-multiplets, the attractor mechanism ensures that its  near-horizon geometry is equivalently described in terms of  $\mathcal{N}=2$  supergravity with just one vector multiplet - the graviphoton, i.e. minimal supergravity in five dimensions. A classification of the solutions of minimal supergravity in five dimensions can be found in \cite{Gauntlett}. This property simplifies the task of finding the relevant Killing spinors for these black holes.  In Section 3 we obtain the near horizon killing spinor in global coordinates and  analyze the $\kappa$-symmetry for stationary probe branes in global time along the lines of \cite{StromingerSB,Strominger5Drot}. In particular we find BPS two-branes wrapped on a holomorphic two-cycle in the Calabi-Yau. These correspond to the zero-branes found in \cite{Strominger5Drot} and have the right properties to be the relevant degrees of freedom for deriving the OSV-relation for black holes with D6-charge. In addition we find BPS five-branes which wrap either a holomorphic four-cycle in the Calabi-Yau and an $S^1$ in space-time or wrap the horizon $S^3/\Z_{p^0}$ completely and a holomorphic two-cycle in the Calabi-Yau. These may play a role analogous to the horizon wrapped two-branes for D4-D0-black holes \cite{StromingerBHQM}. We plan to report on these issuses in subsequent work. 

\section{Near Horizon Geometry of D6-D4-D2-D0 Black Holes}
In order to be self-contained and to fix the conventions we first review the relevant  static half BPS solutions of four-dimensional $\mathcal{N}=2$ supergravity with $n_v$ vector multiplets \cite{FKS}.  The general stationary BPS configurations were derived in \cite{BLS,CdeWKM, Denef}. We then describe the lift of these solutions to five dimensions \cite{Strominger4D5D} and determine the near horizon geometry for a given set of four-dimensional charges.

We consider static single-centered BPS solutions in four dimensions. These solutions are characterized by their asymptotic magnetic and electric charges $(p^I,q_I)$, $I=0,\dots,n_v$ and their asymptotic  moduli. As such they are completely determined in terms of $2n_v +2$ real harmonic functions on ${\mathbbm{R}}^3$
\begin{equation}
H^I(r) = h^I + \frac{p^I}{r}\,, \qquad H_I(r) = h_I + \frac{q_I}{r}\,, \label{harmonic}
\end{equation}
subjected to the condition 
\begin{equation}
p^I h_I - q_I h^I = 0\,. \label{BPS2}
\end{equation}
The corresponding metric is given by
\begin{equation}
\ud s^2_{(4)} = - \frac{\pi}{S(r)} \ud t ^2 + \frac{S(r)}{\pi} \ud
\vec{x}^2\,.
\end{equation}
The function $S$ can be expressed as
\begin{equation}
S = 2 \pi \sqrt{H^0 Q^3 - (H^0 L)^2} \, , \label{S}
\end{equation}
with
\begin{eqnarray}
L &=& \frac{H_0}{2} + \frac{H^A H_A}{2 H^0} + \frac{C_{ABC} H^A H^B H^C}{6
(H^0)^2} \, ,\nonumber \\
Q^{3/2} &=& \frac{1}{6} C_{ABC} y^A y^B y^C \, . 
\end{eqnarray}
Here $A,B,C \in \{1,\ldots,n_v\}$ and $y^A$ are implicitly determined by the equation
\begin{eqnarray}
C_{ABC} y^B y^C  = 2 H_A + \frac{C_{ABC} H^B H^C}{H^0}\,.
\end{eqnarray}

The gauge potentials are determined again by the harmonic functions (\ref{harmonic}) and $S(r)$ (\ref{S})
\begin{equation}
A^I_{(4)} = \frac{1}{S} \frac{\p S}{\p H_I} \ud t  + \mathcal{A}^I\,, \qquad \ud \mathcal{A}^I = *_3 \ud H^I\,.
\end{equation}

To complete the four-dimensional description of generic D6-D4-D2-D0 attractor black holes we give the complex scalar fields
\begin{eqnarray}
t^A = \frac{H^A + \frac{\ui}{\pi} \frac{\p S}{\p H_A}} {H^0 +
\frac{\ui}{\pi} \frac{\p S}{\p H_0}}\,.
\end{eqnarray}

As mentioned above for generic values of these charges the string coupling becomes large in the near horizon regime. To allow for a unified description for generic charges we now give the lift of these solutions to five dimensions. For a nice discussion of this lift see for example \cite{Cheng}. 

The five dimensional metric is given by
\begin{eqnarray}
\ud s^2_{(5)} &=& 2^{2/3} \mathcal{V}^2 (\ud \psi + A^0_{(4)})^2 +
2^{-1/3} \mathcal{V}^{-1} \ud s^2_{(4)} \\
&=& -(2^{2/3} Q)^{-2} (\ud t  + 2 L (\ud \psi + \mathcal{A}^0))^2 +
(2^{2/3} Q) (\frac{1}{H^0} (\ud \psi + \mathcal{A}^0)^2 + H^0 \ud\vec{x}^2)\, , \nonumber
\end{eqnarray}
with
\begin{eqnarray}
\mathcal{V} = \left( \frac{1}{6} C_{ABC} \Im t^A \Im t^B \Im t^C\right)^{1/3} = \frac{S}{2\pi H^0 Q}\,.
\end{eqnarray}
The four and five dimensional gauge potentials are related by 
\begin{eqnarray}
A^A_{(5)} &=& \Re t^A (\ud \psi + A^0_{(4)}) - A^A_{(4)} \nonumber \\
&=& - \frac{Y^A}{2 Q} \ud t + \left( \frac{H^A}{H^0} - \frac{L}{Q} Y^A\right) (\ud \psi + \mathcal{A}^0) - \mathcal{A}^A
\end{eqnarray}
where we introduced the five dimensional scalars
\begin{eqnarray}
Y^A = \frac{y^A}{Q^{1/2}}\,.
\end{eqnarray}
They obey the relation
\begin{eqnarray}
\frac{1}{6} C_{ABC} Y^A Y^B Y^C = 1\,.
\end{eqnarray}

Let us now take the near horizon limit, $r \ll \{p^I,q_I\}$, of the five-dimensional solution.  For this we define 
\begin{eqnarray}
\sigma &=& \frac{1}{r}\,, \nonumber \\
\ud \Omega_3^2 &=& \ud \theta^2 + \sin^2\theta \ud \phi^2 + (\ud \psi/p^0 + \cos\theta  \ud \phi)^2\,, \nonumber \\
R^2_{\text{AdS}} &=& \lim_{r\rightarrow 0} \left(2^{2/3} H^0 Q r^2 \right)\,,  \\
J &=& \lim_{r \rightarrow 0} \left(\frac{(H^0)^{1/2} L}{Q^{3/2}} \right) \qquad \text{and} \nonumber \\
Y^A_0 &=& \lim_{r \rightarrow 0} Y^A\,. \nonumber
\end{eqnarray}
Then, rescaling $t$ appropriately  (denoted again by $t$) we obtain \cite{Ansar}
\begin{eqnarray}
\ud s^2_{(5)} &=& R^2_{\text{AdS}} \left(- \left( \frac{\ud t}{\sigma} + J (\ud \psi/p^0 + \cos\theta \ud \phi)\right)^2 + \frac{\ud \sigma^2}{\sigma^2} + \ud \Omega_3^2 \right) \,, \label{near6} \\
A^A_{(5)} &=& - \frac{Y^A_0 2^{2/3}}{\sqrt{3}} A + \frac{p^A}{p^0} \ud \psi \,,\\
A &=& \frac{\sqrt{3}}{2} R_{\text{AdS}} \left( \frac{\ud t}{\sigma} + J(\ud \psi /p^0 + \cos\theta \ud \phi)\right) \,. \label{near9}
\end{eqnarray}

The near horizon geometry depends on three parameters: the D6 charge $p^0$, the $\text{AdS}_2$ radius $R_{\text{AdS}}$ which is determined by the value of $Q$ at the horizon and the five-dimensional angular momentum $J$. These  are also the quantities that appear in the Beckenstein-Hawking entropy.  In other words, as pointed out in \cite{Strominger4D5D} (see also \cite{Cardoso}) the Taub-NUT fibration, which interpolates between the four-dimensional and the five-dimensional geometry gives a simple geometric representation  of the entropy formula for the 4D-entropy of D6-D4-D2-D0 black holes based on special geometry  \cite{Shmakova}. For $J \rightarrow 1$ a closed light-like curve develops. This singular limit has recently been analyzed in \cite{Strominger07}. 

\section{Global Coordinates and Half BPS-Branes} 
Next we introduce global coordinates. For this we start with the 
expressions (\ref{near6}) and (\ref{near9}) for the metric and 
gauge field respectively and change coordinates as ($\sin B := J$)
\begin{eqnarray}
t&=&\frac{\cos B\cosh\chi \sin\tau}{\cosh\chi \cos\tau+\sinh\chi }\,, \nonumber \\
\sigma&=&\frac{1}{\cosh\chi \cos\tau+\sinh\chi}\,,\\
\psi^{poinc}&=&\psi+2\tan B \tanh^{-1}(e^{-\chi}\tan(\frac{\tau}{2}))\,.\nonumber \label{global1}
\end{eqnarray}
The metric and field strength of the graviphoton then take the form
\begin{eqnarray}
\ud s^2_{(5)} &=& R^2_{\text{AdS}}\left(-\cosh^2\chi \ud \tau^2 +\ud\chi^2+(\sin B\sinh\chi\ud\tau-\cos B\sigma_3)^2+\ud \Omega_2^2 \right)\,,\nonumber\\
F&=&\frac{\sqrt{3}}{2} R_{\text{AdS}}\left(\cos B\cosh\chi\ud\chi\wedge\ud\tau-\sin B\sin\theta\ud\theta\wedge\ud\phi\right)\,,\label{global2} 
\end{eqnarray}
where $\ud \Omega_2^2$ is the line element of the unit two-sphere and 
\begin{eqnarray}
\sigma_3&=&\ud \psi /p^0 + \cos\theta \ud \phi\,. \nonumber
\end{eqnarray}

It is straightforward to lift this near horizon geometry to eleven dimensions. The lifted geometry is a direct product of the five-dimensional space and a Calabi-Yau three-fold $X$ with K\"ahler form $Y^A_0 \omega_A $ where $\omega_A \in H^2(X,\Z)$ is a basis. The three-form $C^{[3]}$ in eleven dimensions is proportional to the wedge product of the gauge field (\ref{near9}) with the K\"ahler form. The Killing spinor equation in eleven dimensions is given by
\begin{eqnarray}
0 &=& \left[\nabla_M+\frac{1}{288}\left(\Gamma_M^{\hphantom{M}N_1N_2N_3N_4} - 8 \delta_M^{N_1} \Gamma^{N_2N_3N_4} \right) G_{N_1N_2N_3N_4} \right] \epsilon \otimes \eta\,, \nonumber \\
\nabla_M &=& \p_M + \frac{1}{4} \omega_{MAB} \Gamma^{AB} \qquad \text{and} \qquad G = \ud C^{[3]}\,, 
\end{eqnarray}
where the capital indices run from zero to ten. The Killing equation on the Calabi-Yau is solved by the covariant constant spinors $\eta_\pm$ on the Calabi-Yau. We have chosen the following conventions: the $\gamma$-matrices on the $X$ w.r.t. an orthonormal frame we denote by $\rho^\textup{\bf i}$ with $\{\rho^\textup{\bf i} , \rho^\textup{\bf j} \} = 2 \delta^{\textup{\bf i} \textup{\bf j}}$. The spinors $\eta_\pm$ obey the relations
\begin{eqnarray}
\rho_{(7)} \eta_\pm = \pm \eta_\pm\,, \nonumber \\
\rho_{\bar i} \eta_+ = 0, \qquad \rho_i \eta_- =0\,,
\end{eqnarray}
where $i$ and $\bar i$ are indices w.r.t. complex coordinates on $X$. The $\gamma$-matrices of five-dimensional space w.r.t. an orthonormal frame we denote by $\gamma^a$ with $\ui \gamma^{01234} = \id$, such that the eleven-dimensional $\gamma$-matrices read $\Gamma^a = \gamma^a \otimes \rho_{(7)}$ and $\Gamma^\textup{\bf i} = \id\otimes  \rho^\textup{\bf i} $.

With a convenient choice of the f\"unf-beine 
\begin{eqnarray}
e^0&=&\cosh\chi\ud\tau,\qquad e^1=\ud\chi\,, \nonumber  \\
e^2&=&\ud\theta,\qquad e^3=\sin\theta\ud\phi\,,\\
e^4&=&\cos B\left(\frac{\ud\psi}{p^0}+\cos\theta\ud\phi\right)-\sin B\sinh\chi\ud\tau\,,\nonumber \label{global3}
\end{eqnarray}
the five-dimensional part of the Killing equation becomes
\begin{eqnarray}
0&=&\partial_\psi\epsilon\,,   \nonumber\\
0&=&\left(\partial_\phi+\frac{1}{2}\cos B\sin\theta\gamma^{24}-\frac{1}{2}\cos\theta\gamma^{23}+\frac{i}{2}\sin B\sin\theta\gamma^2\right)\epsilon\,, \nonumber  \\
0&=&\left(\partial_\theta-\frac{1}{2}\cos B\gamma^{34}-\frac{i}{2}\sin B\gamma^3\right)\epsilon\,,  \\
0&=&\left(\partial_\tau+\frac{1}{2}\sin B\cosh\chi\gamma^{14}-\frac{1}{2}\sinh\chi\gamma^{01}-\frac{i}{2}\cosh\chi\cos B\gamma^1\right)\epsilon\,, \nonumber \\
0&=&\left(\partial_\chi-\frac{1}{2}\sin B\gamma^{04}+\frac{i}{2}\cos B\gamma^0\right)\epsilon\,.\label{global6}  \nonumber
\end{eqnarray}
These equations are solved by (see also \cite{Strominger5Drot})
\begin{eqnarray}
\epsilon&=&S(B,\chi,\tau,\theta,\phi)\epsilon_0\,,  \nonumber \\
S(B,\chi,\tau,\theta,\phi)&=&e^{-\frac{i}{2}B\gamma^4}e^{-\frac{i}{2}\chi\gamma^0}e^{\frac{i}{2}\tau\gamma^1}e^{\frac{1}{2}\theta\gamma^{34}}e^{\frac{1}{2}\phi\gamma^{23}}\label{global7}
\end{eqnarray}
and $\epsilon_0$ is an arbitrary, constant four-component spinor. 

In the following we classify the stationary supersymmetric probe branes in this background. This implies in particular that they are static in $\text{AdS}_2$, i.e. $\dot{\chi}=0$ but allows for M2-branes orbiting around the three-dimensional horizon as well as M5-branes partially or fully wrapping the horizon.

\subsection{Half BPS M2-Branes in Global Coordinates}
We begin with the set of stationary, supersymmetric M2-branes wrapping a holomorphic two-cycle in $X$. The $\kappa$-symmetry condition is \cite{Becker}
\begin{eqnarray}
\Gamma \; \epsilon\otimes\eta &=&\epsilon\otimes\eta \label{hBPSg3}
\end{eqnarray}  
with
\begin{eqnarray}
\Gamma&=&\frac{1}{(p+1)!\sqrt{\det h}}\epsilon^{\hat a\hat b\hat c}\Gamma_{\hat a\hat b\hat c}\nonumber\\
&=&\frac{1}{\sqrt{h_{00}}}\frac{\ud X^\mu}{\ud\tau}e^a_\mu\gamma_a\otimes \ui\id\,,
\end{eqnarray}
where we have assumed that the M2-brane has positive orientation. The hatted indices are the world-volume coordinates and $h_{\hat a \hat b}$ is the pull-back of the space-time metric to the world-volume. 
The second line is expressed in static gauge $\dot X^0=1$.

Let us first consider the case where the two-brane sits at fixed $\theta$ and $\phi$ but rotates in the $\phi$ direction. For $\dot\phi\neq \pm 1$ the BPS condition can never be satisfied. For $\dot\phi=1$ we have $\sqrt{|h_{00}|}=|\cos(B)\sinh(\chi)+\sin(B)\cos(\theta)|$ and 
\begin{eqnarray}
\Gamma_{(0)}:= \frac{\ud X^\mu}{\ud\tau}e^a_\mu\gamma_a&=&-\cosh\chi \gamma^0+\sin\theta \gamma^3-\sin B \sinh\chi \gamma^4 +\cos B \cos\theta \gamma^{4}
\end{eqnarray}
and
\begin{eqnarray}
S^{-1}\Gamma_{(0)}S&=&\cos B \cosh\chi \cos\tau (-\gamma^{0}+\gamma^{4})+i(\cos B \sinh\chi+\sin B \cos\theta)\gamma^{04}\nonumber \\
&&+i\cos B \cosh\chi \sin\tau(\gamma^{10}-\gamma^{14})-i\sin B \sin\theta(\cos\phi \gamma^{03}-\sin\phi \gamma^{02})\nonumber \\ 
&&+i\sin B \sin\theta(\cos\phi \gamma^{43}-\sin\phi \gamma^{42})\,. \label{p1}
\end{eqnarray}
Requiring the $\kappa$-symmetry condition be independent of $\tau$ implies\begin{equation}
\gamma^{04}\epsilon_0=-\epsilon_0\,. \label{t0}
\end{equation}
Furthermore if the latter condition is fullfilled we have 
\begin{eqnarray}
\Gamma\epsilon\otimes\eta&=&\left(\frac{i\Gamma_{(0)}}{\sqrt{h_{00}}}\epsilon\right)\otimes\eta=\epsilon\otimes\eta
\end{eqnarray}
which is just the BPS condition (\ref{hBPSg3}). These solutions correspond to the zero branes found in  \cite{Strominger5Drot}. 
Note that this brane is also BPS for $\dot\phi=1$ while sitting at the north 
pole $\theta=0$ on the base $S^2$.  
This does not mean that this brane is static. 
Indeed, as the velocity along the fiber is given by $\dot\psi/p^0+\cos\theta\dot\phi$, this configuration is geometrically equivalent to that with 
$\dot\phi=0$ and $\dot\psi=p^0$ which is a trajectory along the fiber of the $S^3/\Z_{p^0}$-bundle, i.e. a ``great circle'' on $S^3/\Z_{p^0}$. If we assume instead that $\phi=const, \theta=const$, then $\dot\psi=p^0$ necessarily and the BPS condition reads
\begin{eqnarray}
[-\cos(\theta)\gamma^{04}+\sin(\theta)(\cos(\phi)\gamma^{03}-\sin(\phi)\gamma^{02})]\epsilon_0&=&\epsilon_0\,.\label{condbr}
\end{eqnarray}
So the M2-brane can move along the fiber with constant velocity $p^0$ and sits at any point of the base space $S^2$. The  condition (\ref{condbr}) reduces to (\ref{t0}) for $\theta=0$. 

For $\dot\psi=0$, $\phi=\phi_0$ ($\phi=\phi_0+\pi$) constant and necessarily $\dot\theta=1$ ($\dot\theta=-1$) the BPS condition becomes
\begin{eqnarray}
(\cos(\phi)\gamma^{20}+\sin(\phi)\gamma^{30})\epsilon_0=\epsilon_0.
\end{eqnarray}
Geometrically, this is the case where the M2-brane moves along a meridian of the base $S^2$ with constant velocity one and does not move along the fiber of the $S^3/\Z_{p^0}$-bundle over $S^2$.

To summarize, an M2-brane on $C_2$ is BPS if and only if it rotates with unit angular velocity on the covering space $S^3$. For $\chi=0$ they describe uncharged null-geodesics on $S^3$, while for $\chi>0$ the $M2$-branes are charged and follow a time-like trajectory\footnote{See \cite{Strominger5Drot} for a detailed analysis of the corresponding Born-Infeld action.}. The rotation is required to stabilize them at fixed $\chi$.  This interpretation is compatible with the four-dimensional analysis in \cite{StromingerSB} where it was observed that existence of static half-BPS branes requires that the symplectic product of the charge vector of the probe brane with the background charges does not vanish. Let us consider rotation along the fiber first. Then, since the above results for the wrapped M2-branes are independent of $B$ we can consider vanishing $B$. In this case the non-vanishing of the symplectic product in four dimensions requires that the two-brane rotates along the fiber. Invoking rotational invariance on $S^3$ we then  conclude that rotation along any geodesic circle of the $S^3$ will lead to a half BPS-state.

Note also that a M2-brane sitting for example at $\theta=0$ and rotating in the fiber
with $\dot\psi=p^0$ preserves the same supersymmetry as an 
anti-M2-brane (i.e. negative orientation) at $\theta=\pi$ and $\dot\psi=p^0$. Thus these branes are mutually BPS. This property makes them natural candidates to extend 
the observation of \cite{StromingerOSV} to black holes with D6-charge.  

\subsection{Half BPS Five-Branes}
We now consider stationary M5-branes which partially, or fully wrap the horizon of the five-dimensional black hole. The remaining dimensions of the five-brane are wrapped on a holomorphic cycle in $X$. 

\subsubsection{M5 on $C_4 \times Y$}
Since the pull back of the RR-field strength, $\ud C^{[3]}$, to the world-volume of the five-brane vanishes, we can consistently set the world-volume three-form field strength $F^{\hat a \hat b \hat c}$ to zero. We will assume that the five-brane is wrapped holomorphically on $C_4$. Then the CY-part of $\Gamma$ is just $-1$ \cite{StromingerSB}, so that 
\begin{eqnarray}
\Gamma&=&\frac{1}{(p+1)!\sqrt{\det h}}\epsilon^{\hat a\hat b\hat c \hat d \hat e\hat f}\Gamma_{\hat a\hat b\hat c\hat d\hat e \hat f}\nonumber\\
&=&-\frac{1}{2\sqrt{h}}\frac{\epsilon^{\hat a\hat b}\partial X^\mu \partial X^\nu }{\partial \sigma^{\hat a}\partial \sigma^{\hat b}}e^a_\mu e^b_\nu\gamma_{ab}\otimes \id\,.\label{hBPS52}
\end{eqnarray}
\\
The generic situation can be understood by distinguishing three different $S^1$-wrappings:
{\it i) $Y=(\tau,\theta)$}: here 
\begin{equation}
\sqrt{h}=\sqrt{\cosh^2(\chi)-\sin^2(B)\sinh^2(\chi)}\,. 
\end{equation}
We then conclude that the brane is BPS ($\Gamma\epsilon =\epsilon$) for $\chi=0$ provided 
\begin{equation}
e^{-\frac{1}{2}\phi\gamma_{23}}\gamma_{02}e^{\frac{1}{2}\phi\gamma_{23}}\epsilon_0=-\epsilon_0\,.
\end{equation}
There is no condition on $B$ and $\psi$. This brane wraps a geodesic circle in the horizon  $S^3/\Z_{p^0}$ and is uncharged w.r.t. background fluxes. \\
{\it ii) $Y=(\tau,\phi)$}: For $B=0$ and  $\chi=0$ the brane is BPS for all values of $\theta$ provided
\begin{equation}
\gamma_{04}\epsilon_0=-\epsilon_0\,.
\end{equation}
In this case the brane wraps a geodesic in $S^3/\Z_{p^0}$ and is uncharged.  For $B\neq 0$ the brane is BPS only for $\chi=0$ and $\theta=\frac{\pi}{2}$ with $\gamma_{04}\epsilon_0=-\epsilon_0$. \\
{\it iii) $Y=(\tau,\psi)$}: here  the brane is BPS for $\chi=0$ and $B=0$ provided 
\begin{equation}
e^{-\frac{1}{2}\phi\gamma_{23}}e^{-\frac{1}{2}\theta\gamma_{34}}\gamma_{04}e^{\frac{1}{2}\theta\gamma_{34}}e^{\frac{1}{2}\phi\gamma_{23}}\epsilon_0=-\epsilon_0.
\end{equation}
There is no solution for $B\neq 0$. 

To summarize, a M5 on $C_4 \times Y$ is BPS provided it wraps a maximal geodesic  circle in the squashed horizon,  $S^3/\Z_{p^0}$. From the ten-dimensional perspective these results may be interpreted as follows: An M5-brane wrapped along the $S^2$ 
base becomes an NS5-brane in ten-dimensions which is clearly uncharged and therefore static at $\chi=0$. If the M5-branes is wrapped along the $S^3$-fiber instead, this will become a $D4$-brane with charge vector aligned with that of the background. This brane cannot be static in global time unless the background flux vanishes, i.e. $B=0$. Note that the absence of static branes wrapped along the fiber does not exclude stationary branes. Indeed for rotation in the $\phi$ direction, $\dot\phi=\pm1$, we have 
\begin{equation}
\Gamma=\frac{-1}{\sqrt{h}}\left[ \cosh\chi \cos B \gamma_{04} \pm \sin\theta \cos B \gamma_{34} \right]
\end{equation} 
with
\begin{equation}
\sqrt{det~h}=\sqrt{\cos^2 B(\sinh^2\chi+\cos^2\theta)}\,. 
\end{equation}
The BPS condition is then given by
\begin{eqnarray}
\gamma^{04} \epsilon_0 &=& \mp \epsilon_0\,,  \nonumber \\
\sinh\chi &=& \mp \tan B \cos\theta\,.
\end{eqnarray}
These solutions correspond to rotating BPS configurations found in  \cite{Strominger5Drot}.

\subsubsection{M5 on $C_2 \times S^3/\Z_{p^0}$}
The induced metric $h_{\hat a \hat b}$ is in this case 
\begin{equation}
\begin{pmatrix}
-\cosh^2(\chi)+\sin^2(B)\sinh^2(\chi)&0&-\sin(B)\sinh(\chi)\cos(B)\cos(\theta)&-\frac{\sin(B)\sinh(\chi)\cos(B)}{p^0}\cr
0&1&0&0\cr
-\sin(B)\sinh(\chi)\cos(B)\cos(\theta)&0&\sin^2(\theta)+\cos^2(\theta)\cos^2(B)&\frac{\cos(\theta)\cos^2(B)}{p^0}\cr
-\frac{\sin(B)\sinh(\chi)\cos(B)}{p^0}&0&\frac{\cos(\theta)\cos^2(B)}{p^0}&\frac{\cos^2(B)}{(p^0)^2}
\end{pmatrix}\nonumber
\end{equation}
and
\begin{equation}
 \sqrt{|h|}=|\cosh(\chi)\cos(B)\sin(\theta)/p^0|.
\end{equation}
We will again assume that the brane is wrapped holomorphically on $C_2$ so that the $CY$-part of $\Gamma$ is $\id\otimes\ui\rho_{(7)}$. 

For $B=0$ we then have
\begin{eqnarray}
\Gamma&=&\frac{1}{4!\sqrt{h}}\frac{\epsilon^{\hat a\hat b\hat c \hat d}\partial X^\mu \partial X^\nu \partial X^\rho\partial X^\delta}{\partial \sigma^{\hat a}\partial \sigma^{\hat b}\partial \sigma^{\hat c}\partial \sigma^{\hat d}}e^a_\mu e^b_\nu e^c_{\rho}e^d_\delta\,\Gamma_{abcd}\,(\id\otimes\ui \rho_{(7)})\label{hBPS62}\\
&=&\ui\gamma_{0234}\otimes\rho_{(7)}\nonumber
\end{eqnarray}
so that the brane is BPS for $\chi=0$ and $ (i\gamma_{0234}\otimes\rho_{(7)})(\epsilon_0\otimes\eta)=\epsilon_0\otimes\eta$. 

Next we consider the possibility of non-vanishing world-volume three-form flux $F^{\hat a \hat b \hat c}$ corresponding to M2-branes wrapping $C_2$ and bound to the M5-brane. For this we write 
\begin{equation}
F=-f\, e^2\wedge e^3\wedge e^4-f\,{}^{*_6}(e^2\wedge e^3\wedge e^4 )\,.\label{sdH}
\end{equation}
Here $f$ is proportinal to the number of two-branes. Note that $e^a$, $a=2,3,4$ are the viel-beine on the unit three-sphere, not the three-sphere with radius $R_{\text{AdS}}$ on which the world volume is wrapped and which determines the induced metric relevant for the ${}^{*_6}$ operation. Thus 
\begin{equation}
{}^{*_6}(e^2\wedge e^3\wedge e^4) =\frac{\text{Vol}(C_2)}{R_{\text{AdS}}^2}e^0\wedge e^5\wedge e^6\,,
\end{equation}
where $e^0$ is as in (\ref{global3}) and $e^5$ and $e^6$ are the zwei-beine on $C_2$ with unit volume. Since in the supergravity approximation $\frac{\text{Vol}(C_2)}{R_{\text{AdS}}^2}<\!<1$ we can neglect the last term in (\ref{sdH}). 

With this in mind we will now analyze the $\kappa$-symmetry condition. We find the representation of \cite{Bao} most convenient. Adapting the corresponding projector $\Gamma$ to our situation we get 
\begin{eqnarray}
\Gamma&=&\frac{1}{\sqrt{1-\frac{1}{4}f^2}}\left(\frac{1}{6!\sqrt{\det h}}\epsilon^{\hat a\hat b\hat c \hat d \hat e\hat f}\Gamma_{\hat a\hat b\hat c\hat d\hat e \hat f}-\frac{1}{2\cdot 3!}F^{\hat a\hat b\hat c}\gamma_{\hat a\hat b\hat c}\right)\nonumber\\
&=&\frac{1}{\sqrt{1-\frac{1}{4}f^2}}\left(i\gamma_{0234}\otimes\rho_{(7)}+\frac{1}{2}f\gamma_{ 2 3 4}\right)\,.
\end{eqnarray}
The BPS condition $\Gamma\epsilon =\epsilon$ then implies that  $|f|=2\tanh(\chi)$ with  
\begin{equation}
\begin{cases}
i\gamma_{0234}\epsilon_0=\epsilon_0\quad\hbox{and}& \rho_{(7)}\eta=\eta\,\qquad\hbox{for}\quad  f>0\cr
i\gamma_{0234}\epsilon_0=-\epsilon_0\quad\hbox{and}& \rho_{(7)}\eta=-\eta\,\,\quad\hbox{for}\quad f<0\,.
\end{cases}
\end{equation}
Upon double dimensional reduction along the fiber of $S^3/\Z_{p^0}$ to ten dimensions we get a D4-brane with
\begin{equation}
F_{\hat a\hat b}= F_{4\hat a\hat b}
\end{equation}
which in turn is SUSY according to \cite{StromingerSB}.

Let us now analyze the case with non-vanishing four-brane flux $B\neq 0$. We make the Ansatz
\begin{equation}
F=g\,e^0\wedge e^2\wedge e^3\,. \label{sdH2}
\end{equation}
The $\kappa$-symmetry projector becomes
\begin{equation}
\Gamma=\frac{1}{\sqrt{1+\frac{1}{4}g^2}}\left(i\gamma_{0234}\otimes\rho_{(7)}-\frac{1}{2} g\gamma_{ 0 2 3}\right)\,.
\end{equation}
The BPS condition then implies $\chi=0$ and 
\begin{equation}
\begin{cases}
i\gamma_{0234}\epsilon_0=\epsilon_0\,,\quad \rho_{(7)}\eta=\eta & \frac{1}{2} g=\tan(B) \,\qquad\hbox{or}\cr
i\gamma_{0234}\epsilon_0=-\epsilon_0\,,\rho_{(7)}\eta=-\eta &\frac{1}{2} g=-\tan(B)\quad \,. \label{sdH3}
\end{cases}
\end{equation}
We note in passing that the value of $g$ is not fixed by the equation of motion since $F$, as in (\ref{sdH2}), is a solution of the homogeneous equation $\text{d}{}^*F=0$. On the other hand the killing spinor depends on $C^{[3]}$ through (\ref{global7}) and thus $g$ is fixed by the $\kappa$-symmetry condition (\ref{sdH3}).  

If both, $f$ and $g$ are non-vanishing the $\kappa$-symmetry projector takes the form
\begin{equation}
\Gamma=\frac{1}{\sqrt{1+\frac{1}{4}(g^2-f^2)}}\left(i\gamma_{0234}\otimes\rho_{(7)}+\frac{1}{2}(f\gamma_{234}- g\gamma_{ 0 2 3})\right)\,.
\end{equation}
and the BPS-condition reads 
\begin{equation}
\begin{cases}
i\gamma_{0234}\epsilon_0=\epsilon_0\,,\qquad\rho_{(7)}\eta=\eta\,,\quad\frac{1}{2} g=\tan(B),&\frac{1}{2}f=\frac{\tanh(\chi)}{\cos B}\,,\qquad\hbox{or}\cr
i\gamma_{0234}\epsilon_0=-\epsilon_0\,,\quad\rho_{(7)}\eta=-\eta\,,\quad\frac{1}{2} g=-\tan(B),&\frac{1}{2}f=-\frac{\tanh(\chi)}{\cos B}\,.
\end{cases}
\end{equation}
Thus a static M5-brane wrapped on the horizon and a two-cycle in $X$ with M2-branes on $C_2$ bound to it is BPS for certain values of $\chi$. 

\section{Conclusions}
We have constructed supersymmetric probe branes, stationary in global coordinates of the eleven-dimensional near-horizon geometry, of a generic four-dimensional, single-centered attractor black hole. The motivation for this study came from the success \cite{StromingerBHQM,StromingerOSV} in approximating the black hole partition function by a dilute gas of non-interacting probe branes  in the near horizon geometry of attractor black holes without D6-charge. Our results should provide the necessary ingredients for extending this approach to include D6-charge as well. In particular,  we expect the M2-branes found here to be relevant for understanding the OSV partition function in the presence of D6-charge. Similarly the horizon wrapping M5-branes should contribute, as collective excitations, to the partition function of the conformal quantum-mechanical system dual to the $\text{AdS}_2$ near horizon geometry. 

\paragraph{Acknowledgments:}
The authors would like to thank 
P. Mayr,
P. Kraus,
G. Cardoso and
G. Policastro
for helpful discussions. This work was supported in parts by SFB-375, SPP-1096 and Transregio-33 of the DFG. J.~D.~L. is supported by  a DFG Fellowship with contract number LA 1979/1-1. I.~S. would like to thank the department of Physics and Astronomy at UCLA and the Physics Department at Caltech for hospitality during the final stage of this work. 

\clearpage


\end{document}